\documentclass[aps,prb,floatfix,twocolumn,superscriptaddress,showpacs,amsmath,amssymb, 10pt]{revtex4}
\usepackage{graphicx}
\usepackage{epsfig} 
\allowdisplaybreaks

\newcommand{\be}{\begin{equation}}
\newcommand{\ee}{\end{equation}}
\newcommand{\bea}{\begin{eqnarray}}
\newcommand{\eea}{\end{eqnarray}}
\newcommand{\ba}{\begin{eqnarray*}}
\newcommand{\ea}{\end{eqnarray*}}
\newcommand{\dagga}{{\phantom{\dagger}}}

\newcommand{\bk}{\mathbf{k}}

\newcommand{\dis}{\displaystyle}

\newcommand{\fract}[2]{\frac{\dis #1}{\dis #2}}
\newcommand{\Tr}{\mathrm{Tr}}
\newcommand{\eqn}[1]{(\ref{#1})}

\begin{document}

\title{Sub-ohmic two-level system representation of the Kondo effect} 
\author{Pier Paolo Baruselli} \email{e-mail: barusell@sissa.it}
\affiliation{International
  School for Advanced Studies (SISSA), and CNR-IOM Democritos, 
  Via Bonomea 265, I-34136 Trieste, Italy} 

\author{Michele Fabrizio} 
\affiliation{International School for
  Advanced Studies (SISSA), and CNR-IOM Democritos, Via Bonomea
  265, I-34136 Trieste, Italy} 
\affiliation{The Abdus Salam
  International Centre for Theoretical Physics (ICTP), P.O.Box 586,
  I-34014 Trieste, Italy}

\date{\today} 


\begin{abstract}
It has been recently shown that the particle-hole symmetric Anderson 
impurity model can be mapped onto a $Z_2$ slave-spin theory without any 
need of additional constraints. Here we prove 
by means of Numerical Renormalization 
Group that the slave-spin behaves in this model like a two-level system 
coupled to a sub-ohmic dissipative environment. It follows that the 
$Z_2$ symmetry gets spontaneously broken at zero temperature, which 
we find can be identified with the on-set of Kondo coherence, being the 
Kondo temperature proportional to the square of the order parameter. 
Since the model is numerically solvable, the results are very enlightening 
on the role of quantum fluctuations beyond mean field in the context of 
slave-boson approaches to correlated electron models, an issue 
that has been attracting interest since the 80's. Finally, our
results suggest as a by-product that the paramagnetic metal phase 
of the Hubbard model 
at half-filling, in infinite coordination lattices and at zero temperature, 
as described for instance by Dynamical Mean Field Theory, 
corresponds to a slave-spin theory with a spontaneous 
breakdown of a local $Z_2$ gauge symmetry.   
           
\end{abstract}
\maketitle


Mott's localization, and all its annexes like the local moment formation and 
the Kondo effect, is a phenomenon that escapes  
any mean-field single-particle description since 
it directly affects only part of the electrons' degrees of freedom, 
namely their charge. 
This is unfortunate because mean-field theory is the simplest and 
straightest way to approach interacting many-body systems. 
A trick to circumvent this difficulty, which has provided lots of physical 
insights along the years, is to artificially enlarge the Hilbert space adding 
new degrees of freedom that aim to describe just the electron charge 
configurations, supplemented by local constraints that project the enlarged 
Hilbert space onto the physical one. The final scope is to make Mott's 
localization accessible already at the mean-field level. The most famous 
realization of this work-programme 
is the so-called {\sl slave boson} technique, 
originally introduced to describe Anderson and Kondo models 
for $f$-electron systems.\cite{Barnes-slave-boson,Coleman-slave-boson} 
A delicate issue of the slave-boson theory is that the mean-field treatment, 
although providing quite satisfactory results, 
explicitly breaks a local $U(1)$ gauge symmetry, which cannot be broken hence requires going beyond 
mean-field to be restored.\cite{Read-quantum-fluctuations,Bickers-RMP,Arrigoni}
More recently, novel approaches have been proposed in the attempt 
of reducing the dimension of the enlarged Hilbert space, 
hence the redundancy of the representation, still maintaining the nice 
feature of making Mott localization accessible 
at the mean-field level.\cite{De-Medici,Z2-1,Z2-2} For instance, in 
Ref.~\onlinecite{Z2-2} it has been shown that it is sufficient to introduce 
additional slave Ising variables on each site, namely just two level systems 
rather than infinite level ones as in the original slave-boson technique, 
to account at the mean-field level for a Mott transition in the half-filled 
Hubbard model. In this new representation the continuous $U(1)$ local 
gauge symmetry of the slave-boson theory is replaced by a discrete $Z_2$ one. 
Alike in the slave-boson theory, the $Z_2$ slave-spin formulation 
must be supplemented by a constraint that projects the enlarged Hilbert space 
onto the physical one. Remarkably, it has been shown in 
Ref.~\onlinecite{Marco-PRB} that the constraint is 
unnecessary in the case of an Anderson impurity model at particle-hole 
symmetry, and, equivalently, of a Hubbard model at 
half-filling in the limit of infinite coordination lattices, which 
can be mapped\cite{DMFT} onto an impurity model self-consistently coupled 
to a bath. 
Specifically, given the Anderson impurity Hamiltonian
\bea
\mathcal{H}_{\text{AIM}} &=& \sum_{\bk\sigma}\, \epsilon_\bk 
c^\dagger_{\bk \sigma}c^\dagga_{\bk\sigma} \nonumber\\
&& + \sum_{\bk\sigma}\, V_\bk \Big(d^\dagger_\sigma c^\dagga_{\bk\sigma}
+ H.c.\Big) + \frac{U}{2}\,\big(n_d-1\big)^2\nonumber\\
&\equiv& \mathcal{H}_{\text{bath}} + \mathcal{H}_{\text{hyb}} + 
\frac{U}{2}\,\big(n_d-1\big)^2,\label{H-AIM}
\eea
where $c^\dagger_{\bk\sigma}$ creates a conduction electron while $d^\dagger_\sigma$ 
an impurity one, with 
$n_d=\sum_\sigma\,d^\dagger_\sigma d^\dagga_\sigma$, 
and the Ising plus electron model
\be
\mathcal{H}_{Z_2} = \mathcal{H}_{\text{bath}} + 
\sigma^x\,\mathcal{H}_{\text{hyb}} + \frac{U}{4}\,\big(1-\sigma^z\big),
\label{H-Z2}
\ee
where $\sigma^a$, $a=x,y,z$, are Pauli matrices, it follows 
that at particle-hole symmetry the following identity 
holds:\cite{Marco-PRB} 
\be
Z_{AIM} \equiv \Tr\bigg(\text{e}^{-\beta\mathcal{H}_{\text{AIM}}}\bigg)
= \frac{1}{2}\,Z_{Z_2} \equiv \frac{1}{2}\,
 \Tr\bigg(\text{e}^{-\beta\mathcal{H}_{Z_2}}\bigg).\label{Z-Z}
 \ee
\begin{figure}[t]
\centering
\includegraphics[width=8cm]{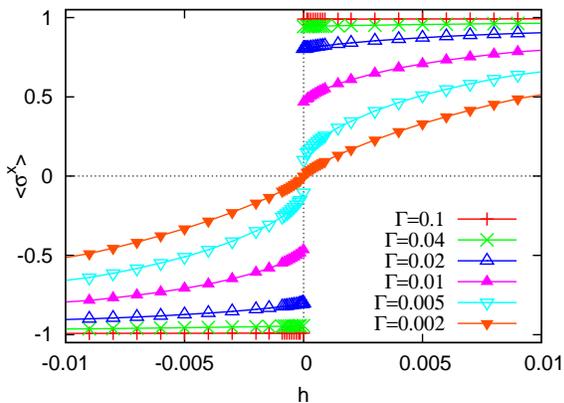}
\caption{The average value $\langle \sigma^x\rangle$ as a function 
of the external symmetry breaking field $h$ for several values of $\Gamma$
and $U=0.1$.}
\label{sigma_x}
\end{figure}
The Ising operator $\sigma^z$ in \eqn{H-Z2} can be identified with 
the electron operator $1-2\left(n_d-1\right)^2$, which has value +1 
if $n_d=1$ and -1 if $n_d=0,2$, hence it describes charge fluctuations.
Furthermore, the mixed operator $\sigma^x d^\dagga_\sigma$ actually 
represents the physical impurity annihilation operator.  
At zero temperature, the two Hamiltonians \eqn{H-AIM} and \eqn{H-Z2} 
must therefore have the same ground state energy. It can be readily shown 
that a mean-field factorized wavefunction 
$\mid \Psi\rangle = \mid \text{Ising}\rangle \times 
\mid \text{electrons}\rangle$ 
for the model \eqn{H-Z2} allows to reproduce all mean-field results of 
the slave-boson mean-field approach to the Anderson impurity 
Hamiltonian \eqn{H-AIM}. In fact, if we assume 
that $\langle \text{Ising}\mid \sigma^z\mid \text{Ising}\rangle = \cos\theta$ 
and $\langle \text{Ising}\mid \sigma^x\mid \text{Ising}\rangle = \sin\theta$, 
then the electronic wavefunction must be the ground state of the resonant 
level Hamiltonian 
\be
\mathcal{H}_* = \mathcal{H}_{\text{bath}} + \sin\theta\,
\mathcal{H}_{\text{hyb}} + \frac{U}{4}\,\big(1-\cos\theta\big),\label{H*}
\ee
with energy $E(\theta)$ and hybridization width 
$\Gamma_* = \sin^2\theta\,\Gamma$, lower than its bare value $\Gamma$. 
Minimization of 
$E(\theta)$ leads to the same result as obtained by slave-boson 
mean-field theory.\cite{Schonhammer} For instance, if we assume that 
$\Gamma(\epsilon)= \Gamma$ for $\epsilon\in [-D,D]$ and zero otherwise, 
assumption that we shall make hereafter taking $D=1$ our unit of energy,
then, for $U\gg \Gamma$, we find that the known mean-field result  
\be
\sin^2\theta = \frac{1}{\Gamma}\,\exp\left(-\fract{\pi U}{16\Gamma}\right).
\label{sintheta-MF}
\ee
By analogy with slave bosons, it is tempting to interpret  
the finite value of $\sin\theta$ as manifestation 
of Kondo coherence, and the impurity operator $d_\sigma$ in 
\eqn{H-Z2} as the coherent quasiparticle with weight 
$\sin^2\theta$. In fact, within mean field the spectral 
function $A_{d\sigma}(\epsilon)$ 
of the physical electron $\sigma^x\,d^\dagga_\sigma$ is simply the 
convolution of the spectral functions of the resonant level, 
$A_d(\epsilon)$ and of the Ising operator $\sigma^x$,
\be 
A_\sigma(\epsilon)= \sin^2\theta\,\delta(\epsilon) 
+\frac{\cos^2\theta}{2}\,\Big(\delta(\epsilon-\Omega) 
+\delta(\epsilon+\Omega)\Big),\label{Asigma-HF}
\ee
where $\Omega= U/2\cos\theta$ is the effective 
Zeeman splitting of the Ising spin.
Specifically, one finds that 
\bea
&& A_{d\sigma}(\epsilon)  \rightarrow \Big(A_d \ast A_\sigma\Big)(\epsilon) =
\sin^2\theta 
A_d(\epsilon) \label{A-HF} \\
&& + \frac{\cos^2\theta}{2}\,\Big(\theta(\epsilon-\Omega)\,
A_d(\epsilon-\Omega)  
+  \theta(-\epsilon-\Omega)\,A_d(\epsilon+\Omega)\Big),\nonumber
\eea
The mean-field expression of  
$A_{d\sigma}(\epsilon)$ displays a low energy resonance with width 
$\Gamma_*$ and weight $\sin^2\theta$, the rest of its weight 
being concentrated in two symmetric peaks 
at energies $\pm \Omega$. It is quite reasonable 
to regard these latter as 
the Hubbard side-bands and instead 
the central peak as the Abrikosov-Suhl resonance, 
hence $\Gamma_*$ the Kondo temperature $T_K$. However, while the mean field 
result predicts that the Hubbard bands are broadened on the same 
scale $\Gamma_*$ as the central resonance, in reality their width 
is controlled by the bare $\Gamma$.   

\begin{figure}[t]
\centering
\includegraphics[width=8cm]{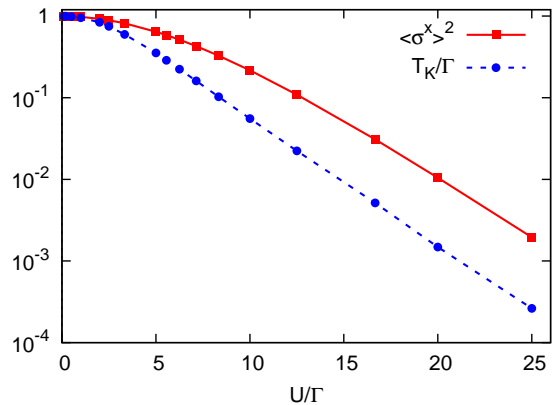}
\caption{Square of the symmetry breaking order parameter and of 
$T_K/\Gamma$ plot in logarithmic scale versus $U/\Gamma$. Note the linear 
behavior at large $U/\Gamma$.}
\label{sigma.vs.TK}
\end{figure}
Another questionable aspect of the mean field 
solution with $\langle \sigma^x \rangle \not = 0$ is that it explicitly 
breaks the discrete symmetry 
of \eqn{H-Z2}: $d_\sigma \to - d_\sigma$ and $\sigma^x\to -\sigma^x$.
Therefore, just like in the conventional slave-boson 
approach,\cite{Read-quantum-fluctuations} 
we may wonder how reliable is mean field. We are going to show that, 
unlike in slave-boson theory, 
the breaking of the discrete $Z_2$ symmetry of \eqn{H-Z2} is not an artifact 
of mean-field but does spontaneously occur in the actual ground state. 

We start noticing that the model \eqn{H-Z2} 
resembles a two-level system coupled to a dissipative 
bath,\cite{Leggett-RMP} where the two levels are the states with 
$\sigma^x=\pm 1$ and the 
tunneling is provided by $\sigma^z$. In reality, the two models are not 
rigorously equivalent, since the hybridization 
operator $\mathcal{H}_{\text{hyb}}$ does not behave exactly like the 
coupling to a dissipative bath of bosons. However, 
if we judge solely from the low frequency behavior of the dissipative-bath 
spectral function, $J(\omega)\sim \omega^s$,\cite{Leggett-RMP}   
we should conclude that the model \eqn{H-Z2} 
behaves like a spin-boson Hamiltonian with an extremely 
sub-ohmic dissipation, $s=0$. 
In such a case, we expect the tunneling to be irrelevant at low 
energy, hence unable to split the degeneracy of the two levels. 
This conclusion is remarkable because it means 
that the mean field result is more correct than we would have expected; 
the $Z_2$ symmetry is spontaneously broken at zero temperature. 
\begin{figure}[t]
\centering
\includegraphics[width=8cm]{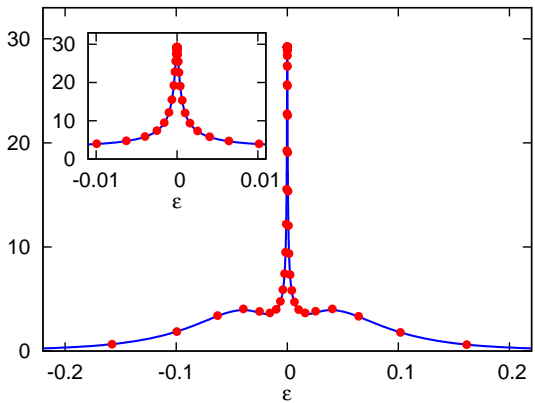}
\caption{Spectral function of the physical impurity electron operator, 
$\sigma^x d_\sigma$, 
at $U=0.1$ and $\Gamma=0.01$.
The dots are the values of the impurity spectral 
function as calculated directly by the Anderson impurity model \eqn{H-AIM}. 
We just plot few of these points, 
since the two spectral functions are just coincident.}
\label{impurity-spectra}
\end{figure}
There is still another aspect that we find worth mentioning. 
In reality, the Anderson impurity model \eqn{H-AIM} con be  
mapped into another spin-boson model\cite{Hamann} 
with the role of the spin being played now by the physical spin 
of the impurity, rather than by the charge as in \eqn{H-Z2}. 
In this alternative and more familiar representation, 
the bath is however ohmic and leads to incoherent delocalization of the 
physical spin, once again the Kondo effect.  
We find quite amusing that 
the Kondo effect, known to occur in the Anderson impurity model 
\eqn{H-AIM} with constant $\Gamma$, happens to be described 
by two complementary spin-boson models, 
in one of which it corresponds to delocalization while in the other, 
the model Eq.~\eqn{H-Z2}, to localization.  

The above speculation can be supported by actual calculations, affordable 
because of the reduced dimension of the Hilbert space as opposed to 
conventional slave bosons. In particular, we shall consider both 
the Anderson impurity model, Eq.~\eqn{H-AIM}, and the model \eqn{H-Z2}, 
and study them by means of the 
Numerical Renormalization Group (NRG).\cite{Wilson_RMP}

The first task we undertake is to confirm 
that spontaneous breaking of $Z_2$ does occur 
in model \eqn{H-Z2}. We hence add to $\mathcal{H}_{Z_2}$ 
a symmetry breaking 
perturbation $-h\,\sigma^x$, and calculate the average 
$\langle \sigma^x \rangle$ as function of $h$. The results are shown in 
Fig.~\ref{sigma_x}, where it is evident the characteristic behavior of a 
symmetry broken phase. The asymptotic value of $\langle \sigma^x \rangle$ 
as $h\to 0^+$ can be considered as the zero-field order parameter. In 
Fig.~\ref{sigma.vs.TK} we plot $\langle \sigma^x \rangle^2$ 
as well the actual Kondo temperature in units of $\Gamma$ 
of the Anderson impurity model \eqn{H-AIM} 
on a logarithmic scale as function of $U/\Gamma$. We showed that 
within mean field these two quantities coincide, 
but in reality they do not, see Fig.~\ref{sigma.vs.TK}, 
even though the linear slope at large $U/\Gamma$ is the same, 
and twice as 
bigger as the mean field value $-\pi U/16\Gamma$, Eq.~\eqn{sintheta-MF}. 
This simply means that mean field as usual 
overestimates the value of the symmetry breaking order parameter, 
hence of $T_K$.  

\begin{figure}[t]
\centering
\includegraphics[width=8cm]{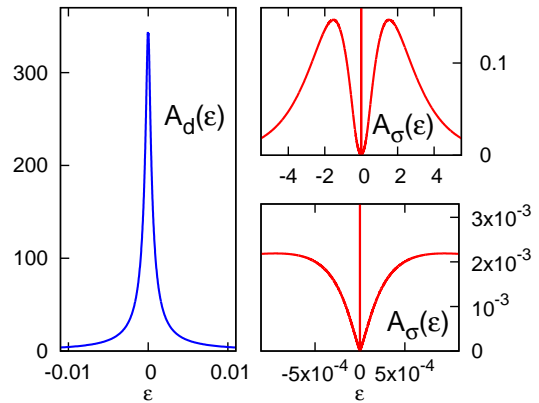}
\caption{Left panel: Spectral function of the impurity operator 
$d_\sigma$. Right panel: Spectral function of the Ising operator 
$\sigma^x$, on a large scale, top, and zoomed close to zero energy, bottom.
The parameter that are used are $U=0.1$ and $\Gamma=0.01$ in units of 
the conduction electron half-bandwidth.}
\label{spectra}
\end{figure}
The next important step is showing that gauge invariant quantities in 
both models \eqn{H-AIM} and \eqn{H-Z2} are indeed the same. 
We then calculate 
the impurity spectral function of the Anderson impurity model and compare 
it with the spectral function of the physical electron 
$\sigma^x\,d^\dagga_\sigma$ in model \eqn{H-Z2}. 
The two quantities are found to be practically 
indistinguishable from each other, see Fig.~\ref{impurity-spectra}, thus 
confirming the validity of the mapping.

We previously showed that within mean field theory, 
the spectral function of the physical electron $A_{d\sigma}(\epsilon)$ 
is approximated by the convolution between the spectral 
functions $A_\sigma(\epsilon)$ 
of $\sigma^x$, whose mean field expression is given in 
Eq.~\eqn{Asigma-HF}, and 
$A_d(\epsilon)$ of $d^\dagga_\sigma$, in mean field simply 
the spectral function of a resonant level 
model of width $\sin^2\theta\,\Gamma$. The actual NRG  
$A_d(\epsilon)$ is found not to differ much from the mean field result; 
it is still made up of a single resonance at the chemical potential, see 
Fig.~\ref{spectra}, of width the true Kondo temperature. 
Instead, $A_\sigma(\epsilon)$ is substantially different from mean field, 
see Fig.~\ref{spectra}. 
It still displays a $\delta$-peak at $\epsilon=0$ with weight 
$\langle \sigma^x \rangle^2$, even though not exactly coincident 
with the value extracted in the zero-field limit, see the comment below. However, the finite energy peaks shift at much 
higher energy, around the edge of the particle-hole continuum, 
and get quite broadened. In addition, a tiny linear in $\epsilon$ component 
appears at low energy, which resembles much the particle-hole spectrum of 
the resonant level since the linear behavior stops just around 
the Kondo temperature. This result shows that quantum 
fluctuations couple strongly the Ising spin with the electrons, 
providing a very short lifetime to the Ising spin-flip excitations. 
Although NRG, at least in the version we use, is not expected to be very accurate at high energy, especially in such a case of a sub-ohmic bath,\cite{Vojta&Bulla} 
we believe that the gross features of $A_\sigma(\epsilon)$ that we find, 
e.g. the very broad incoherent background peaked at high energy, are true.

\begin{figure}[t]
\centering
\includegraphics[width=8cm]{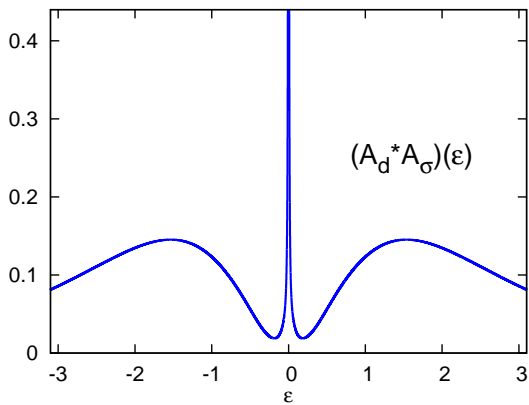}
\caption{Convolution of the spectral functions $A_\sigma(\epsilon)$ 
and $A_d(\epsilon)$ for $U=0.1$ and $\Gamma=0.01$. 
Note the absence of Hubbard bands, which should be appear at 
$\sim \pm U/2=0.05$ but are hidden below the 
broad background peaked around the particle-hole band edge at energy 
$\sim \pm 2D = \pm 2$.}
\label{comparison}
\end{figure}

We observe that the spectral function $A_{d\sigma}(\epsilon)$ 
of the physical electron $\sigma^x d^\dagga_\sigma$, shown in 
Fig.~\ref{impurity-spectra}, can be generally written as  
\be
A_{d\sigma}(\epsilon) = \int d\omega\, A_d(\epsilon-\omega)\,
A_\sigma(\omega)\,K(\omega,\epsilon),\label{true}
\ee
where the kernel $K(\omega,\epsilon)$ amounts for all vertex corrections. 
In Fig.~\ref{comparison} we show 
the simple convolution of $A_d(\epsilon)$ 
and $A_\sigma(\omega)$, which, compared with 
the correct result in Fig.~\ref{impurity-spectra}, 
could provide a rough estimate of vertex corrections. 
We notice that, while the Abrikosov-Suhl resonance is reproduced quite well 
also by the convolution, the Hubbard side-bands are completely masked 
by the broad background of $A_\sigma(\epsilon)$, see Fig.~\ref{spectra}. 
This implies that vertex 
corrections play a major role at high energy, filtering out that 
high-energy background and letting the Hubbard bands emerge.

In conclusion, we have shown several amusing features of the $Z_2$ 
slave-spin representation of the 
particle-hole symmetric Anderson impurity model that, 
unlike its slave-boson analogous, can be exactly 
solved by NRG. Specifically, we have shown that in this language the 
Kondo effect corresponds to spontaneous breaking of a local discrete 
$Z_2$ symmetry, that does survive quantum fluctuations.

We end by mentioning that, as a byproduct, our result suggests 
that the zero-temperature paramagnetic 
metal phase of the half-filled Hubbard model, 
in the limit of infinite lattice-coordination,\cite{DMFT} 
can be regarded within the $Z_2$ slave-spin 
representation\cite{Z2-2,Marco-PRB} as a phase where 
a local $Z_2$ gauge symmetry is spontaneously broken. Conversely, the 
zero-temperature Mott metal-to-insulator transition would correspond to the 
restoration of the $Z_2$ symmetry. We note that this result 
does not violate Elitzur's theorem,\cite{Elitzur} 
because of the infinite lattice-coordination 
limit.\cite{Maslanka} However, we expect that 
the local $Z_2$ gauge symmetry should be recovered at any 
finite temperature, since  
a sub-ohmic dissipative two-level system is known to delocalize at 
any finite temperature.\cite{Leggett-RMP} 

This work has been supported by PRIN/COFIN
20087NX9Y7. We are grateful to Marco Schir\'o for his useful comments.  
   


\end{document}